\documentclass[twocolumn,aps,prl,showpacs,superscriptaddress,
amssymb]{revtex4}

\usepackage{color}
\usepackage{graphicx}
\usepackage{dcolumn}
\usepackage{bm}
\usepackage{textcomp, pifont}

\begin{document}


\title{Superconducting specific heat jump $\Delta C_{\rm el} \propto T_c^\beta \ (\beta \approx 2)$ for K$_{1-x}$Na$_x$Fe$_2$As$_2$}
\author{V.\ Grinenko} \email{v.grinenko@ifw-dresden.de}
\affiliation{Leibniz-Institute for Solid State and Materials Research, IFW-Dresden, D-01171 Dresden, Germany}
\author{D.V.\ Efremov}
\affiliation{Leibniz-Institute for Solid State and Materials Research, IFW-Dresden, D-01171 Dresden, Germany}
\author{S.-L.\ Drechsler} \email{s.l.drechsler@ifw-dresden.de}
\affiliation{Leibniz-Institute for Solid State and Materials Research, IFW-Dresden, D-01171 Dresden, Germany}
\author{S.\ Aswartham}\affiliation{Leibniz-Institute for Solid State and Materials Research, IFW-Dresden, D-01171 Dresden, Germany}
\author{D. Gruner}\affiliation{Leibniz-Institute for Solid State and Materials Research, IFW-Dresden, D-01171 Dresden, Germany}
\author{M.\ Roslova}
\affiliation{Leibniz-Institute for Solid State and Materials Research, IFW-Dresden, D-01171 Dresden, Germany}
\affiliation{
Lomonosov Moscow State University, GSP-1, Leninskie Gory, Moscow, 119991, Russian Federation}
\author{I.\ Morozov}
\affiliation{Leibniz-Institute for Solid State and Materials Research, IFW-Dresden, D-01171 Dresden, Germany}
\affiliation{
Lomonosov Moscow State University, GSP-1, Leninskie Gory, Moscow, 119991, Russian Federation}
\author{K.\ Nenkov}
\affiliation{Leibniz-Institute for Solid State and Materials Research, IFW-Dresden, D-01171 Dresden, Germany}
\author{S.\ Wurmehl}\affiliation{Leibniz-Institute for Solid State and Materials Research, IFW-Dresden, D-01171 Dresden, Germany}
\affiliation{
Institut f\"ur Festk\"orperphysik, TU Dresden,
D-01062 Dresden, Germany}
\author{A.U.B.\ Wolter}\affiliation{Leibniz-Institute for Solid State and Materials Research, IFW-Dresden, D-01171 Dresden, Germany}
\author{B.\ Holzapfel}
\affiliation{Leibniz-Institute for Solid State and Materials Research, IFW-Dresden, D-01171 Dresden, Germany}
\affiliation{Karlsruhe Institute of Technology (KIT),
Hermann-von-Helmholtz-Platz 1, 76344 Eggenstein-Leopoldshafen, Germany}
\author{B.\ B\"uchner}\affiliation{Leibniz-Institute for Solid State and Materials Research, IFW-Dresden, D-01171 Dresden, Germany}
\affiliation{
Institut f\"ur Festk\"orperphysik, TU Dresden,
D-01062 Dresden, Germany}

\date{\today}

\begin{abstract}
We present a systematic study of the electronic specific heat jump ($\Delta C_{\rm el}$) at 
the superconducting transition temperature $T_c$ of  K$_{1-x}$Na$_x$Fe$_2$As$_2$. Both $T_c$ and $\Delta C_{\rm el}$ monotonously decrease with increasing $x$. 
The specific heat jump scales approximately with a power-law, $\Delta C_{\rm el} \propto T_c^{\beta}$, with
$\beta \approx 2$ determined 
by the impurity scattering rate, 
in contrast to most iron-pnictide superconductors, 
where the remarkable
Bud'ko-Ni-Canfield (BNC) scaling $\Delta C_{\rm el} \propto T^3$ 
has been found.
Both the $T$ dependence of $C_{\rm el}(T)$ in the superconducting state and the 
nearly quadratic scaling of 
$\Delta C_{\rm el}$ at $T_c$ are well described by the Eliashberg-theory for a 
two-band $d$-wave superconductor with weak pair-breaking due 
to nonmagnetic impurities. 
The disorder induced by the Na substitution significantly suppresses the small gaps leading to gapless states in the slightly disordered superconductor,
which results in a large observed residual Sommerfeld coefficient in the superconducting state for $x > 0$.
\end{abstract}

\pacs{74.25.Bt, 74.25.Dw, 74.25.Jb, 65.40.Ba}

%
\maketitle 

The overwhelming majority of iron-pnictide superconductors exhibit
several puzzling universal features. One of them is the
Bud'ko-Ni-Canfield (BNC) scaling of the specific heat (SH) jump
($\Delta C_{\rm el}$) at the superconducting transition temperature ($T_c$)
\cite{Budko2009}. For example, Ba(Fe$_{1-x}$Co$_x$)$_2$As$_2$,
Ba(Fe$_{1-x}$Ni$_x$)$_2$As$_2$, and many other compounds
\cite{Budko2009, Kogan2010,Chaparro2012} show a
$\Delta C_{\rm el}\propto T_c^3$ behavior. Several
scenarios were proposed to explain this unusual behavior
\cite{Zaanen2009, Kogan2009, Kogan2010, Kuzmanovski2014}
or at least deviations from the BCS-theory
prediction were ascribed to the interplay of coexisting
spin density wave (SDW) state
and superconductivity (SC) 
\cite{Vavilov2011,Vavilov2011a}.
One of the possible reasons for the
$\Delta C_{\rm el}\propto T_c^3$ behavior might be 
a strong pair-breaking, which 
is an intrinsic property of many Fe pnictide superconductors
\cite{Kogan2009, Kogan2010} due to the vicinity of competing
magnetic (spin-density-wave) phases and/or the always
present impurities. Another approach to explain the cubic BNC scaling
rests on the
assumption of a non-Fermi liquid behavior near a 
magnetic critical point. In such a special situation 
a $T^3_c (T^2_c)$ scaling in three (and two) dimensions, respectively,
based on sophisticated field-theory arguments was
suggested by J.\ Zaanen \cite{Zaanen2009,She2011}. Also, the 
possible influence of thermal SDW fluctuations on the SH jump value was 
emphasized in 
Ref.\ \onlinecite{Kuzmanovski2014}.
Recently, it was found that the ,,universal'' cubic BNC
scaling fails for the heavily
hole-doped K$_x$Ba$_{1-x}$Fe$_2$As$_2$ at a K doping $x>0.7$
\cite{Budko2013}. The authors suggested that the observed
deviations point to significant changes in
the nature of the superconducting state. For example, in
stoichiometric KFe$_{2}$As$_{2}$ $d$-wave \cite{Hashimoto2010, Reid2012,
Abdel-Hafiez2013} or $s$-wave SC with accidental
nodes \cite{Okazaki2012, Watanabe2013, Hardy2013a} were suggested by different
experiments. From the theoretical side it was predicted that with
increasing hole-doping the superconducting order parameter changes
from nodeless $s_{\pm}$ at the optimal doping to a nodal $s_{\pm}$
\cite{Maiti2012} or $d$-wave state \cite{Thomale2011} in KFe$_2$As$_2$ via 
possible intermediate $s+is$ or $s+id$-wave 
states \cite {Maiti2013}, respectively.
In this paper we show that among the Fe pnictides at 
least two
different groups can be distinguished by their $\Delta C_{\rm el}$ vs.
$T_c$ plot. The first group \cite{Budko2009,Budko2013} is related to
the majority of Fe pnictides as proposed previously.
K$_{1-x}$Na$_{x}$Fe$_{2}$As$_{2}$ 
belongs
to the second group which scales almost conventionally
with $\Delta C_{\rm el}\propto T_c^{2}$. 
Both the observed  $\Delta C_{\rm el}$ vs.\  
$T_c$ behavior and the 
$T$ dependence of $C_{\rm el}$ are consistent with multiband 
$d$-wave SC in K$_{1-x}$Na$_{x}$Fe$_{2}$As$_{2}$.

K$_{1-x}$Na$_{x}$Fe$_{2}$As$_{2}$ single crystals 
with a typical mass of about 1-2 mg and several mm in-plane 
dimensions were grown by the self-flux method.
The compositions and the phase purity of the investigated
samples were determined by an EDX analysis in a scanning
electron microscope, and by x-ray analysis 
\cite{supplement}. 
The SH data  were obtained by
a relaxation technique in a Physical Property Measurement System
(PPMS), Quantum Design. The resistivity was measured by 
the standard four-contact method in the PPMS. 
The magnetic dc susceptibility as a
function of temperature was measured using a commercial SQUID
magnetometer, Quantum Design.

\begin{figure}[b]
\includegraphics[width=21pc,clip]{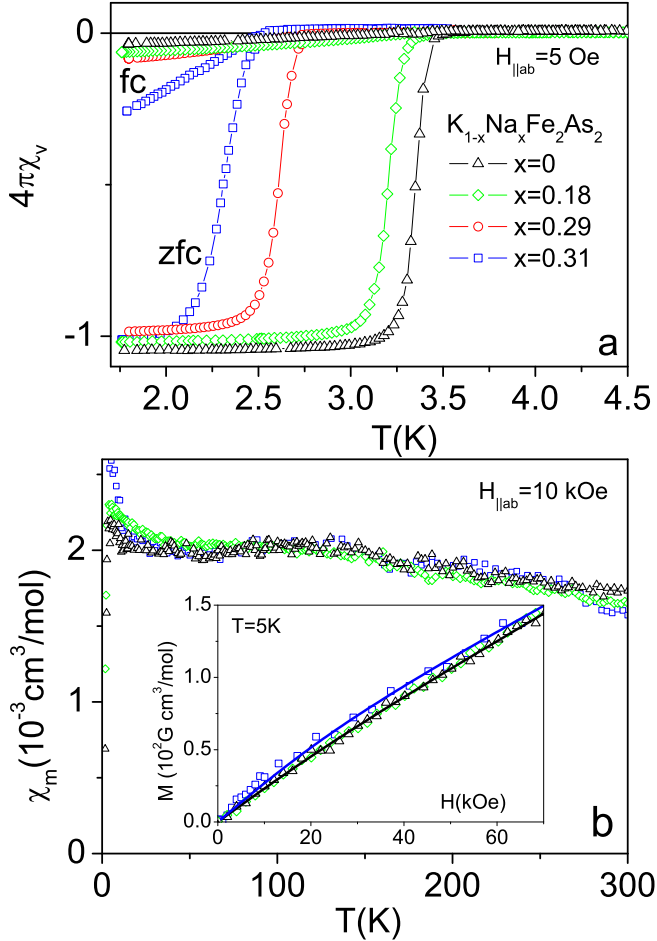}
\caption{(Color online) (a) Temperature dependence of the volume
susceptibility $\chi_v$ measured in a magnetic field of $H
\parallel ab$ = 5\,Oe. (b) Temperature dependence of the molar
susceptibility $ \chi_m$ measured in a magnetic field of $H
\parallel ab$ = 10\,kOe. Inset: magnetization curves measured in an external 
field $H \parallel ab$. 
For explanation of the fitting curves see 
\cite{supplement}.} 
\label{Fig:1}
\end{figure}
\begin{figure}[t]
\includegraphics[width=21pc,clip]{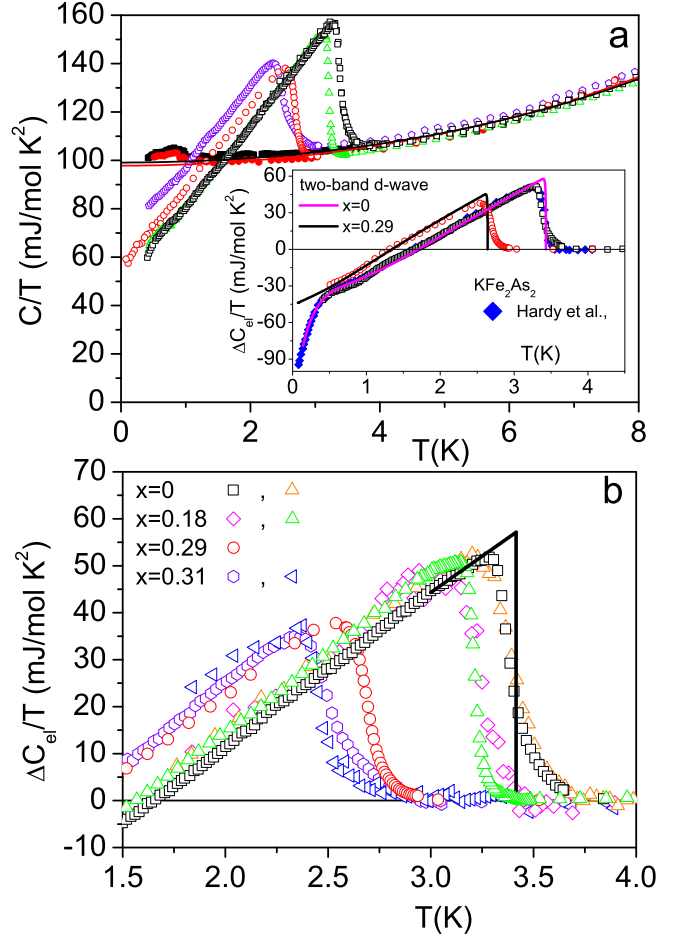}
\caption{(Color online) (a) Total specific heat $C/T$ of
K$_{1-x}$Na$_{x}$Fe$_{2}$As$_{2}$ vs.\ $T$. Open symbols: zero field data, closed symbols: data in an applied magnetic field $H\parallel c$ =
15 kOe, solid lines: the fit of the normal-state specific heat (see text). 
The experimental data for $x$=0.29 at $T <$ 0.4 K are taken from
Ref. \onlinecite{Abdel-Hafiez2013}. Inset: the
specific heat $\Delta C_{\rm el}/T = [C(0)-C(H)]/T$ vs.\ $T$
for a sample with $x$= 0 and 0.29.
Dots: experimental data; filled 
rhombus: data from Ref. \cite {Hardy2013a} shown for comparison; solid
line: calculation within two-band Eliashberg-theory for $d$-wave
superconductors with nonmagnetic impurities and 
$T_{c0}$ = 3.44\,K
for the adopted clean limit. (b) $\Delta C_{\rm el}/T$ near $T_c$ for 
different samples
of K$_{1-x}$Na$_{x}$Fe$_{2}$As$_{2}$. Solid lines are
plotted according to the entropy equal-area construction
method for determining
$T_c$ and $\Delta C_{\rm el}/T_c$.} \label{Fig:2}
\end{figure}

The $T$ dependencies of the volume susceptibility $\chi_v$ of
K$_{1-x}$Na$_{x}$Fe$_{2}$As$_{2}$ for different $x$ values are
shown in Fig.~\ref{Fig:1}a. All investigated samples 
have a large superconducting volume fraction.
The $T$ dependencies of the molar susceptibility $\chi_m$ of
K$_{1-x}$Na$_{x}$Fe$_{2}$As$_{2}$ for different $x$ values measured for
$H\parallel ab$ = 10\,kOe are shown in Fig.~\ref{Fig:1}b. The
$\chi_m$ in the normal state is nearly independent on the
temperature for 50 K $\leq T \leq$ 150 K. This behavior is expected for the 
paramagnetic Pauli
susceptibility of a Fermi liquid. The upturn 
below $T$ = 50 K
indicates a small amount of magnetic impurities 
\cite{Hardy2013}. The measured $\chi_m$ values are considerably 
lower than the data reported previously in Ref.\ 
\onlinecite{Grinenko2013} for KFe$_{2}$As$_{2}$ with a cluster
glass behavior and similar to the Pauli susceptibility value
reported in Ref.\ \onlinecite{Hardy2013}, where an impurity
contribution had already been subtracted. 
The magnetization curves measured at $T$ = 5\,K 
(see the inset in
Fig.~\ref{Fig:1}b) slightly deviate
from a linear dependence. 
This allows us to estimate the
concentration of magnetic impurities $n$ in the 
samples assuming 
that it is related to paramagnetic 
Fe atoms. In this case, we arrived at
$n \lesssim 0.1$ mol\% for all investigated 
samples \cite{supplement}. 
The low value of $n$ suggests that the Na substitution 
does not induce local magnetic moments and that it 
can be considered as a nonmagnetic impurity.
The corresponding linear static susceptibility at low temperatures after 
subtraction of the impurity contribution
$\chi_s\approx 1.8(3)\cdotp 10^{-3}$cm$^3$/mol is independent of the
Na concentration within the error bars of the sample masses. 

The SH $C(T)$ of
K$_{1-x}$Na$_{x}$Fe$_{2}$As$_{2}$ is shown in Fig.\ \ref{Fig:2}. A
clear superconducting anomaly is observed for all $x$ in line with the
susceptibility measurements.
The normal state SH below $T$= 10 K can be fitted by using the standard 
expression $C(T)=\gamma_{n}T + \beta T^{3} + \eta T^{5}$,
where $\gamma_n$ is 
the normal-state Sommerfeld coefficient and 
the next two terms with
$\beta$ and $\eta$ describe 
the lattice contribution to the SH. A value of 
$\gamma_n \approx$ 100 mJ/mol K$^{2}$ was found for all investigated samples
irrespective of the Na 
substitution within the error bars of the sample masses 
\cite {supplement}. This value is in accord with other published 
data for KFe$_2$As$_2$
\cite{Abdel-Hafiez2012,Hardy2013,Fukazawa2011,Kim2011,Budko2012}.
Inspecting Fig.\ \ref{Fig:2}a we see that the SH
in the applied field of
$H \parallel c$ = 15 kOe deviates from a
standard fit at $T\lesssim$ 1.5 K 
forming a small hump, where the
upper critical field obeys
$H_{c2}^{\parallel c}(0) \lessapprox$ 15 kOe for 
K$_{1-x}$Na$_{x}$Fe$_{2}$As$_{2}$ according to
Refs.\  
\onlinecite {Abdel-Hafiez2012,Abdel-Hafiez2013, Hardy2013a, Kittaka2014}.
The complete suppression of the SC by 
$H \parallel c$ = 15 kOe is also 
supported by the resistivity measurements \cite{supplement}.    
Thus, this hump cannot be ascribed to SC.    
A low-$T$ anomaly related to magnetic impurities has been 
reported recently for KFe$_2$As$_2$ in Ref.\ 
\onlinecite{Kim2011}. In our case the entropy 
 confined in this anomaly is about 0.05\% of $R\ln 2$ 
which is comparable to the value 
estimated in Ref.\  \onlinecite{Kim2011}. 
By analogy, the hump can 
be related to a small amount of magnetic impurities
observed in our samples in the magnetization measurements in Fig.\ \ref{Fig:1}. 
However, we cannot exclude 
intrinsic origins of this hump. 
We note that the low-$T$ upturn in the normal state 
was also observed recently in KFe$_2$As$_2$ single crystals 
in the literature
\cite {Hardy2013, Hardy2013a,Kittaka2014}. Therefore, further 
investigations are needed to clarify the nature of this anomalous behavior at 
low $T$. 
Above $T \sim $ 1.5 K the 
electronic SH contains a dominant linear-in-$T$ term 
as expected in the case of a clean Fermi liquid. 
Near $T_c$ the corresponding Wilson 
ratio $R_W = \pi^2 k_B^2\chi/(3\mu_B^2\gamma_n) \sim $ 1 for all investigated 
samples. This  together with the doping independent Sommerfeld coefficient 
indicates
that neither the density of states nor the strength of correlation effects are
affected by the Na substitution. 

The SH contribution related to the SC 
$\Delta C_{\rm el}(T)/T = [C(0,T)-C(H,T)]/T$ is shown in 
Fig.\ \ref{Fig:2}a (inset) and in Fig.\ \ref{Fig:2}b, where $C(0,T)$ is the SH in
zero magnetic field and $C(H,T)$ is the one measured in a magnetic
field $H > H_{c2}$. The latter is
taken to be the SH of the normal state. It is seen in Fig.\
\ref{Fig:2}b that the SH jump $\Delta C_{\rm el}/T_c$ at $T_c$ is a
monotonic function of $T_c$. 
\begin{figure}
\includegraphics[width=20pc,clip]{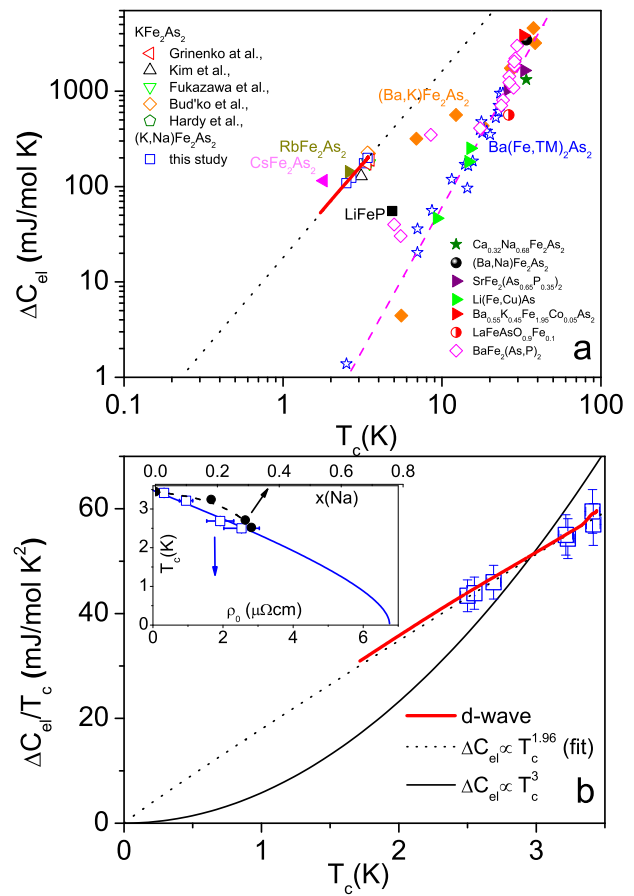}
\caption{(Color online) (a) The dependence of $\Delta C_{\rm el}$ vs.\ $T_c$ for
different  iron-pnictide superconductors. Solid curve: obtained within
Eliashberg theory, dotted curve: the best fit of our experimental data by $\Delta C_{\rm el} \propto T_c^\beta$ and dashed curve:
BNC scaling $\Delta C_{\rm el} \propto T_c^3$ \cite{Budko2009,Budko2013}.
The data are taken from
Refs.\ \onlinecite{Grinenko2013,Kim2011,Fukazawa2011,Budko2012,Hardy2013}
for KFe$_2$As$_2$, CsFe$_2$As$_2$ \cite{Wang2013}, RbFe$_2$As$_2$ 
\ \cite{Stewart2011}, (Ba,K)Fe$_2$As$_2$ and Ba(Fe,TM)$_2$As$_2$ \cite{Budko2013} (TM - transition metal),
LiFeP \cite{Kim2013}, Ca$_{0.32}$Na$_{0.68}$Fe$_2$As$_2$ 
\cite{Johnston2013}, (Ba,Na)Fe$_2$As$_2$ \cite{Aswartham2012},
 SrFe$_2$(As$_{0.65}$P$_{0.35}$)$_2$ \cite{Kobayashi2013}, Li(Fe,Cu)As
\cite{Kim2012, Stockert2011},
Ba$_{0.55}$K$_{0.45}$Fe$_{1.95}$Co$_{0.05}$As$_2$ \cite{Gofryk2012},
 LaFeAsO$_{0.9}$Fe$_{0.1}$ \cite{supplement, Grinenko2011},
BaFe$_2$(As$_{0.7}$P$_{0.3}$)$_2$ \cite{Kim2011a}.
(b) $\Delta C_{\rm el}/T_c$ vs.\ $T_c$ for (K,Na)Fe$_2$As$_2$.
Inset: $T_{\rm c}$ vs. residual resistivity $\rho_0$ 
and the
Na substitution level $x$. Solid line: fit using 
the Abrikosov-Gor'kov formula \cite{supplement}, and the
dashed curve: guide to the eye.}
\label{Fig:3}
\end{figure}
The $\Delta C_{\rm el}$ data of our
K$_{1-x}$Na$_{x}$Fe$_{2}$As$_{2}$ single crystals, and those
taken from the literature for KFe$_{2}$As$_{2}$ samples with
various $T_{\rm c}$, together with many other 
Fe-arsenides and also 
some Fe-phosphides \cite{Walmsley2013,Chaparro2012,
Kim2013,Kim2011a,Hashimoto2010} 
are summarized in Fig.\ \ref{Fig:3}a. 
Two classes can be identified in the $\Delta C_{\rm el}$ vs. $T_c$
plot: The largest group with $\Delta C_{\rm el}\propto T_c^3$, first
reported in Ref.\ \onlinecite{Budko2009,Chaparro2012}, includes the
overwhelming majority of Fe pnictides. Most of the
superconductors in this region are believed to exhibit an
$s_{\pm}$ order parameter. Noteworthy, stoichiometric LiFeAs
\cite{Kim2012, Stockert2011, Morozov2010}, which has 
a relatively low residual resistivity value 
$\rho_0 \sim 1.3 \mu\Omega\cdotp$cm \cite{Rullier-Albenque2012} as 
compared to most of the other Fe-based 
superconductors, and the impure Cu doped
derivative \cite{Kim2012}, also fit to the BNC
scaling, perfectly. 
A clearly distinct second class, we report here, consists of the (K,Na)Fe$_2$As$_2$ 
systems. Phenomenologically this group scales approximately 
as $T_c^{\beta}$ with an exponent $\beta \approx 2$
\cite{remark2,remark4} and
can be associated with the pair-breaking
dependence of $\Delta C_{\rm el}$ in KFe$_2$As$_2$ due to disorder
appearing during the sample synthesis of the stoichiometric compound
or induced by the Na substitution. 
The sister compounds 
RbFe$_2$As$_2$ \cite{Stewart2011} and 
CsFe$_2$As$_2$ \cite{Wang2013} exhibit similar values of the SH jump and of
$T_c$. However, further experimental studies on samples with a
different amount of disorder are necessary to determine whether these 
systems behave similarly to KFe$_2$As$_2$.
Finally, several systems 
do not fit to any of these two
classes: among them are Ba$_{1-x}$K$_{x}$Fe$_{2}$As$_{2}$
\cite{Budko2013} with $0.7<x<1$, LiFeP \cite{Kim2013}, and 
BaFe$_2$(As$_{1-x}$P$_x$)$_2$ for $x \approx 0.2$ and 0.65 \cite{Walmsley2013}. 
(Note that for $0.3 \lesssim x \lesssim 0.4$, in the 
vicinity of a critical point, a $T_c^{6.5 \pm 0.7}$ behavior has been 
suggested in Ref.\  \onlinecite{Walmsley2013}, although the deviations from the 
BNC scaling is hardly visible on the logarithmic scale of Fig.\ \ref{Fig:3}a).

An inspection of Fig.\ \ref{Fig:2}a 
shows that the $C(T)/T$ for $x=0.29$ and 0.31 
is nearly linear  below $T_c$.
For $x=0$ 
the experimental data show a downturn at 
low $T$ which can be 
explained by the presence of a small superconducting gap/gaps. 
Similar conclusions 
were drawn by the authors of Ref. 
\onlinecite{Hardy2013, Hardy2013a, Kittaka2014}.
This low-$T$ superconducting anomaly was not observed down to 0.1K 
for $x$ = 0.29.
The downturn is also absent in 
those KFe$_2$As$_2$ single crystals showing disorder related
magnetic contributions,
\cite{Grinenko2013, Abdel-Hafiez2013} with only slightly reduced $T_c$. 
The enhancement of the impurity scattering with Na 
substitution is evidenced by the 
significant
increase of the residual resistivity 
$\rho_0$  from 
0.32(6) $\mu\Omega\cdotp$cm at $x$=0 to 2.5(5) $\mu\Omega\cdotp$cm at $x$=0.31, 
where the RRR=$\rho(300{\rm K})$/$\rho_0$ value 
strongly decreases almost by an order of
magnitude, from 1080 to 150 \cite{supplement}.
Therefore, by analogy with CeCoIn$_5$ \cite{Barzykin2007} the observed 
unusual behavior of the SH might be ascribed to 
gapless SC in those
bands with small 
superconducting gaps ($\sim \Delta_2$) which is further
suppressed by the disorder induced pair-breaking. (Note, that non-magnetic 
impurities 
also suppress SC in the case of $s_{\pm}$ order parameter \cite{Efremov11}.) 
In 
this case
the Sommerfeld coefficient ($\gamma_2$) at low $T$ 
for the Fermi surface sheets (FSS) with $\Delta_2$  
approaches almost the 
normal-state value, whereas the $\gamma_1(T)$ corresponding to the other FSS 
with the large gap $\Delta_1$ shows 
the behavior expected for a
single-band superconductor but with a formally
large residual SH
value $\gamma_2 \sim \gamma_r \gtrsim 50$ mJ/mol-K$^2$ \cite{remark3}. 
In particular, the temperature dependence of 
$\gamma_1(T) \sim \Delta C_{\rm el}(T)/T \propto T$ at 
low $T$ for 
samples with a part of the quasiparticles in a
gapless state evidences the 
line nodes on the dominant order parameter with
 $\Delta_1$ \cite{Abdel-Hafiez2013}. However, for the 
general description of the SH in clean and dirty samples, strictly speaking,
a detailed investigation of 
a four-band model with a significant number of new parameters would be 
requested. But we believe that for the present level of understanding, 
the study of a less-complex effective two-band model as a minimum model is 
necessary.  

The experimental data of the 
$T$ dependence of the SH (inset of Fig.\ \ref{Fig:2}a) and the SH jump at $T_c$ 
shown in Fig.\ \ref{Fig:3} at all $x$ can be 
reasonably well described by the two-band Eliashberg theory
or a fully nodal $d_{\rm x^2-y^2}$-wave
(in the notations of the folded Brillouin zone with 2 Fe per unit cell) 
superconductor with a weak pair-breaking included, 
but also with two rather differently coupled
two groups of quasi-particles, therefore having rather different
gap values. 
This situation is different 
from the standard  $s_{\pm}$
case considered, e.g.\ in the
optimally-doped (Ba,K)Fe$_2$As$_2$ \cite{Popovich2010}
where usually at least {\it two} strongly interacting
bands dominate the interband-coupling-driven SC
within a higher-order multiband model, also
including further weakly coupled bands.
Note that other symmetries of the superconducting order 
parameter, 
such as $s_{\pm}$ with accidental nodes and $d_{\rm xy}$-wave have been 
proposed 
in the literature for KFe$_2$As$_2$ based on the low-$T$ SH data 
\cite{Hardy2013a,Kittaka2014}. However, the discussion of 
these alternative scenarios is 
beyond the scope of the present paper.

For the calculations in the frame of the Eliashberg theory we 
used the same spin-fluctuation spectra as in 
Ref.\ \onlinecite{Abdel-Hafiez2013} peaked at $\approx 8$ meV and 
an Einstein phonon 
peaked at 20meV, with the
phenomenologically fitted coupling constants in 
two $d$-wave channels: 
$\lambda _{\rm d1} = 0.818$,  $\lambda _{\rm d2} = 0.05$, 
$\lambda _{\rm d12} = \lambda _{\rm d21}N_2/N_1=0.13$ and 
also a weak uniform electron-phonon 
coupling  $\lambda _{\rm s1} = 0.1$,  $\lambda _{\rm s2} = 0.1$, 
$\lambda _{\rm s12} = \lambda _{\rm s21}N_2/N_1=0.04$, with the partial density 
of states of the effective two-band system obeying $N_2/N_1 = 2$. The
nonmagnetic impurity scattering caused by the
smaller, isovalent Na$^+$ ions 
has been treated
in the adopted Born approximation \cite{Preosti1994}. 
For the calculations we adopted  
$\Gamma_2/\Gamma_1$=5, where $\Gamma_1$ and $\Gamma_2$ are the impurity 
scattering 
rates in band 1 and 2, correspondingly. Qualitatively, the Na position
out of the FeAs-block  
position might be
the reason for the different $\Gamma$ values. Therefore, a stronger scattering 
effect for the 
Fe 3$d_{xz}$ and 3$d_{yz}$ orbitals  oriented out of the Fe-plane 
as compared to the
Fe 3$d_{xy}$ orbitals 
might be expected. The value of the critical temperature in 
the clean limit 
adopted for the calculations is $T_{\rm c0}$=3.44K. Note that the $T_c$ 
suppression 
approximately follows the Abrikosov-Gor'kov pair-breaking
theory with 
$\Gamma_{\rm eff} \propto \rho_0$ (see the inset in Fig.\ \ref{Fig:3}b and 
Ref.\onlinecite{supplement}). Thus, adopting the Born approximation, we obtain 
a quantitative phenomenological
description of our data 
shown in Figs.\ \ref{Fig:2} and \ref{Fig:3} within 
an effective two-band Eliashberg theory 
for a nodal
$d$-wave superconductor.

In summary, we have shown that the Fe pnictide superconductors
can be divided at least into two groups according to their
$\Delta C_{\rm el}$ vs. $T_c$ plots. The main group 
contains the overwhelming majority of Fe pnictides.
The second group 
consist of the heavily
hole-doped superconductors (K,Na)Fe$_2$As$_2$ and stands
out from the other Fe pnictides with respect to its absolute
values and its distinct scaling: 
$\Delta C_{\rm el} \propto T_c^{\beta}$ with $\beta \approx 2$.
This behavior and the $T$ dependence of $\Delta C_{\rm el}(T)$ in the
superconducting state are well described by two-band 
Eliashberg-theory for 
$d$-wave superconductors with weak pair-breaking due
to nonmagnetic impurities. 

This work was supported by the DFG through the SPP 1458, the
E.-Noether program (WU 595/3-1 (S.W.)), and the EU-Japan project
(No.\ 283204 SUPER-IRON). S.W.\ thanks the
BMBF for support in the frame of the ERA.Net RUS project FeSuCo
No.\ 245. We acknowledge fruitful discussion with K.\ Iida, 
A.\ Chubukov, S.\ Johnston, O.\ Dolgov, D.\ Evtushinsky 
as well as P.C.\ Canfield J.\ Zaanen, and P.\ Walmsley
for stimulating interest. We 
thank also
P.\ Chekhonin, and E.\ Ahrens for support 
in the experimental performance and valuable discussions.

\newpage
\maketitle

\section*{SUPPLEMENTARY MATERIAL}

In the present 
supplementary part we present our unpublished data for the 
specific heat of LaFeAsO$_{0.9}$F$_{0.1}$ employed in Fig.\ 3a for
the BNC-scaling shown
in the main text. We also provide additional 
data for the 
electrical resistivity of K$_{1-x}$Na$_x$Fe$_2$As$_2$ used 
there in the inset of
Fig.3b. Furthermore, we give a detailed analysis of the magnetic 
impurity contribution in our samples and present the 
fitting parameters of the normal state 
SH. Finally, we provide x-ray data for the 
 investigated single crystals.

\renewcommand{\theequation}{S\arabic{equation}}
\renewcommand{\thefigure}{S\arabic{figure}}
\renewcommand{\thetable}{S\arabic{table}}
\setcounter{equation}{0}
\setcounter{figure}{0}
\setcounter{table}{0}

\section*{The specific-heat of LaFeAsO$_{0.9}$F$_{0.1}$}
\begin{figure}
\includegraphics[width=16pc,clip]{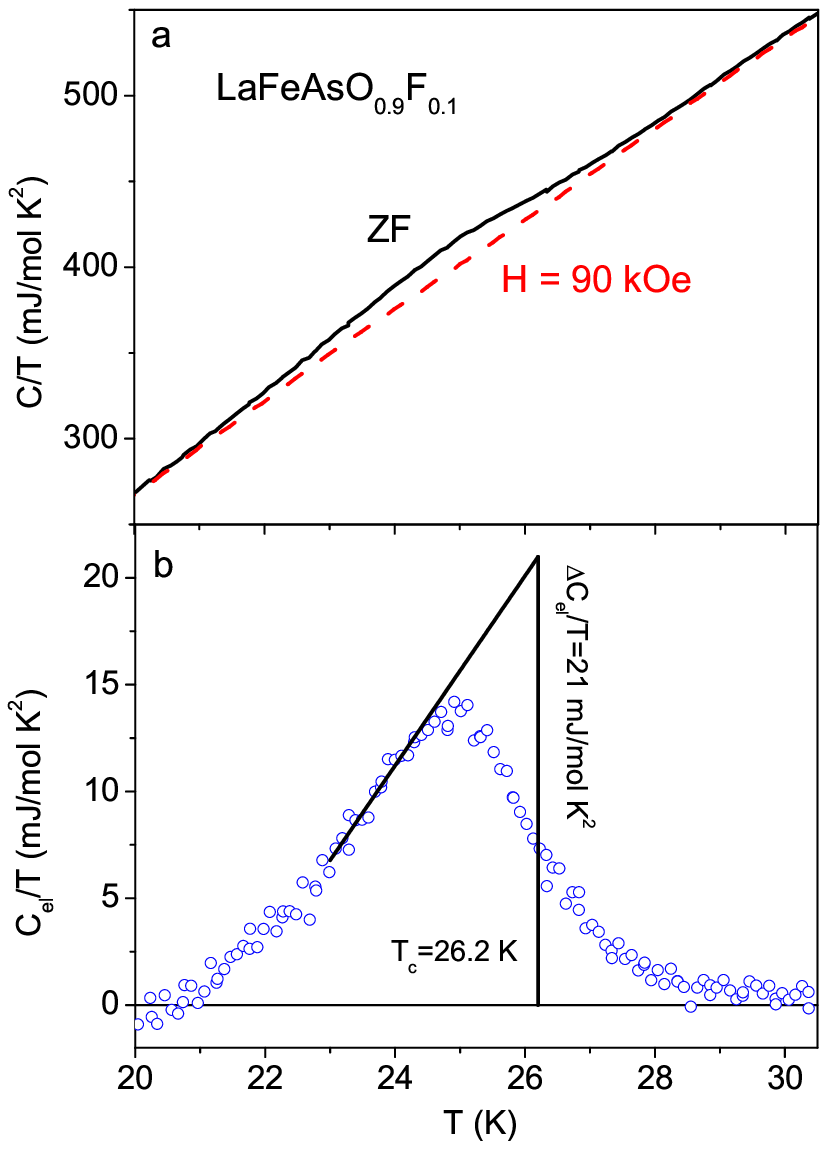}
\caption{(a) Temperature dependence of the specific heat of
LaFeAsO$_{0.9}$F$_{0.1}$ in zero magnetic field (ZF) and
applied magnetic fields $H=$90~Oe. (b) Temperature 
dependence of the electronic specific heat 
$\Delta C_{\rm el}/T = [C(0)-C(9~ \mbox{\rm T})]/T$ 
near $T_c$.}
\label{SH}
\end{figure}

Here, we present our unpublished LaFeAsO$_{0.9}$F$_{0.1}$ 
data with the aim to demonstrate that the 
La-1111 superconductors
join also the cubic BNC-scaling.
Polycrystalline LaFeAsO$_{0.9}$F$_{0.1}$ samples were 
prepared
from pure components using a two-step solid-state reaction method 
\cite{Kondrat2009}. Structural data and other physical properties 
of these samples are given in Refs.\ \onlinecite{Fuchs2009,Grinenko2011}.
In Fig.\ \ref{SH} we show the
specific heat $C$ data for LaFeAsO$_{0.9}$F$_{0.1}$ in 
zero magnetic field (ZF) and applied magnetic fields 
of  H = 90~kOe.
To get an electronic contribution $\Delta C_{\rm el}$ in the superconducting 
state we 
considered the difference $\Delta C_{\rm el} = C(0)-C(9 \mbox{\rm T})$. 
This procedure provides 
$\Delta C_{\rm el}$ in the range of 3 K below $T_{\rm c}$ taking into account 
that 
$dH_{\rm c2}/{T_{\rm c}}\approx 2.85$ T/K \cite{Grinenko2011}. As shown in 
Fig.\ \ref{SH} this temperature range is sufficient to obtain the 
electronic specific heat jump
$\Delta C_{\rm el}$ for this compound.   

\section*{Resistivity data of our K$_{1-x}$Na$_x$Fe$_2$As$_2$ samples}
\begin{figure}
\includegraphics[width=18pc,clip]{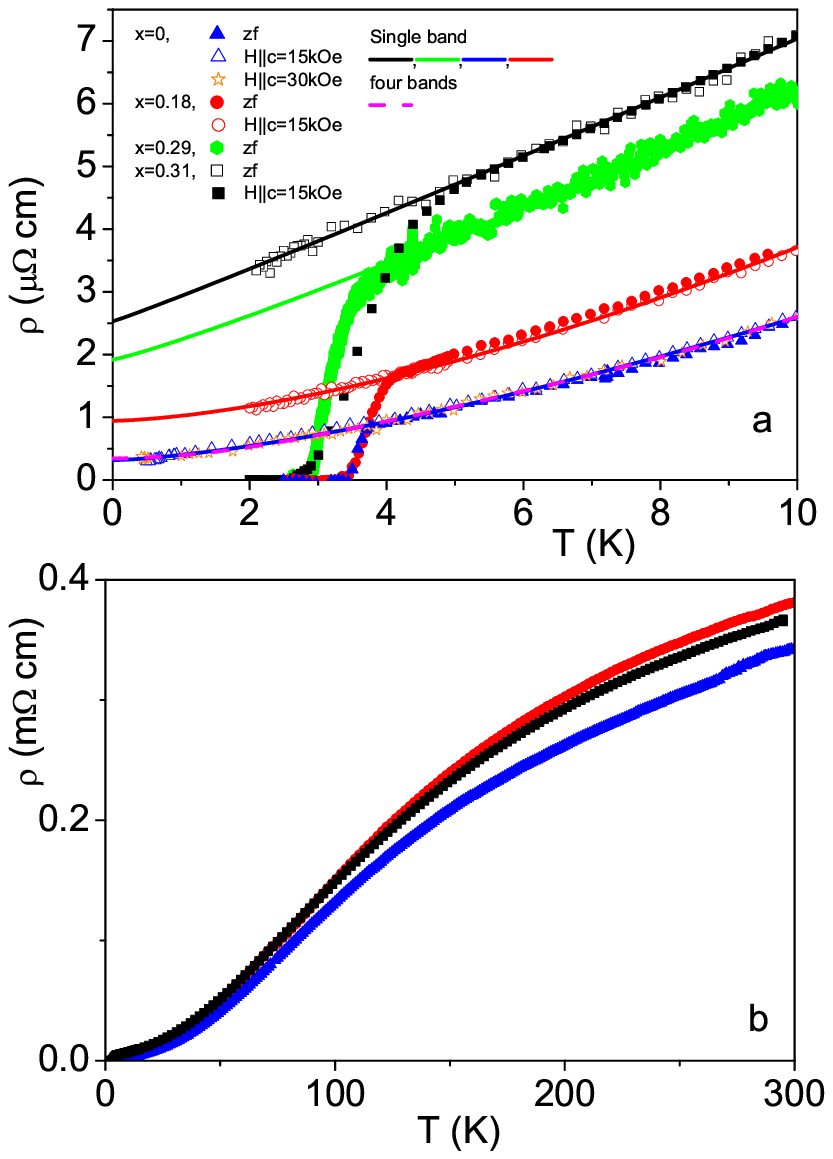}
\caption{
Temperature dependence of the resistivity (a):
below 10~K (the data for $x=0.29$ are taken from 
Ref.\ \onlinecite{Abdel-Hafiez2013}), (b): zero field data up to 300K.}
\label{res}
\end{figure}

The resistivity data of K$_{1-x}$Na$_x$Fe$_2$As$_2$ 
single crystals for various $x$ are 
shown in Fig.\ \ref{res}. In order to 
determine the clean limit value of $T_{\rm c0}$ we plot 
the $T_{\rm c}$ values
of the investigated samples versus the experimental 
values of the residual 
resistivity $\rho_0$ as shown in the inset of 
Fig.\ 3b (main text).
We note that K$_{1-x}$Na$_x$Fe$_2$As$_2$ 
is a multiband system. However, $T_c$ is defined mainly by 
the strong superconducting band. Therefore, for the sake of simplicity, to 
estimate $T_{\rm c0}$ we fitted the experimental data by the single-band
Abrikosov-Gor'kov (AG)
formula modified for the $d$-wave case \cite{Radtke1993,remark}
\begin{equation}
-\ln \left( \frac{T_c}{T_{c0}} \right) =\psi \left( \frac{1}{2}+\frac{\alpha T_{c0}}{2\pi T_c} \right)-\psi \left( \frac{1}{2}\right) \ ,
\end{equation}
where $\alpha = 1/[2\tau_{\rm eff}T_{c0}$] is the strong-coupling pair-breaking
parameter and $\Gamma_{\rm eff}=1/\tau_{\rm eff} \propto \rho_0$ is the 
an effective scattering rate due
to impurities (created by the Na substitution). Then 
the clean limit value  
$T_{\rm c0} = 3.5(1)$~K was obtained according to the analysis of the $T_c$
suppression using $\tau_{\rm eff}$ as an additional fitting parameter.
It is seen in Fig.\ \ref{res}a that below 10~K the temperature dependence of 
resistivity deviates from 
the standard $\rho(T)=\rho_0+AT^{2}$ Fermi-liquid behavior. 
This deviation is stronger for crystals with higher $\rho_0$ values.
We note that this deviation does not necessarily mean that 
our samples are in a non-Fermi liquid regime. In particular, 
a subquadratic dependence can be expected in the case of multiband metals 
even when all the $i$ bands follow the standard law  
 $\rho_i(T)=\rho_{i0}+A_iT^{2}$. In this case the fit by the
multiband equation $\rho(T)=1/\sum_i(1/\rho_i(T))$ ($i$=4) provides 
a good description of the data as shown in Fig.\ \ref{res}a. However, one 
should assume that the scattering 
rates and the Fermi velocities are strongly band dependent. 

\section*{Analysis of the magnetic impurity contribution}

The magnetization data given in the inset of 
Fig.\ 1b of the main text can be fitted using two 
contributions: 
$m(H,T) = m_{int}(H,T)+m_{imp}(H,T)$, where $m_{int}$ is 
the intrinsic magnetization and $m_{imp}$ is 
the magnetic contribution of the 
impurity. We assume that 
$m_{int}(H,T) \approx \chi_s (T)$H is dominated by 
the static spin susceptibility $\chi_s$ and other contributions 
are neglible \cite{Grinenko2011}. To estimate 
the concentration of 
the magnetic impurities $n$ in the 
samples we assumed that the impurity contribution is related to paramagnetic 
Fe$^{+3}$ with $J$=5/2. In 
this case $m_{imp} = nJgB(H/T)$, where $B(H/T)$ is the
Brillouin function, and $g$ is the
Land\' e  factor. From the fit of the experimental 
data shown in the inset of Fig.\ 1b in the
main text, we estimated that $n \lesssim 0.1{\rm mol}\%$ for all investigated 
samples.

\section*{The main parameters for the normal state SH}

The fitting parameters obtained from the fit of the 
normal-state specific heat data by 
$C(T)=\gamma_{n}T + \beta T^{3} + \eta T^{5}$ shown in 
Fig.\ 2a in the main text are:   

For $x$  =0: $\gamma_n$ = 99(2) mJ/mol K$^{2}$, $\beta$ = 0.51(5) mJ/molK$^{4}$, $\eta = 1.2(2)\cdotp 10^{-3}$ mJ/molK$^{6}$;

$x$ = 0.19: $\gamma_n$ = 98(3) mJ/mol K$^{2}$, $\beta$ = 0.51(5) mJ/molK$^{4}$, $\eta = 0.9(2)\cdotp 10^{-3}$ mJ/molK$^{6}$;

$x$ = 0.29: 
$\gamma_n$ = 99(2) mJ/mol K$^{2}$, $\beta$ = 0.51(4) mJ/molK$^{4}$, $\eta = 0.9(2)\cdotp 10^{-3}$ mJ/molK$^{6}$;

$x$ = 0.31: 
$\gamma_n$ = 98(6) mJ/mol K$^{2}$, $\beta$ = 0.52(5) mJ/molK$^{4}$, $\eta = 1.4(4)\cdotp 10^{-3}$ mJ/molK$^{6}$.

\begin{figure}
\includegraphics[width=18pc] {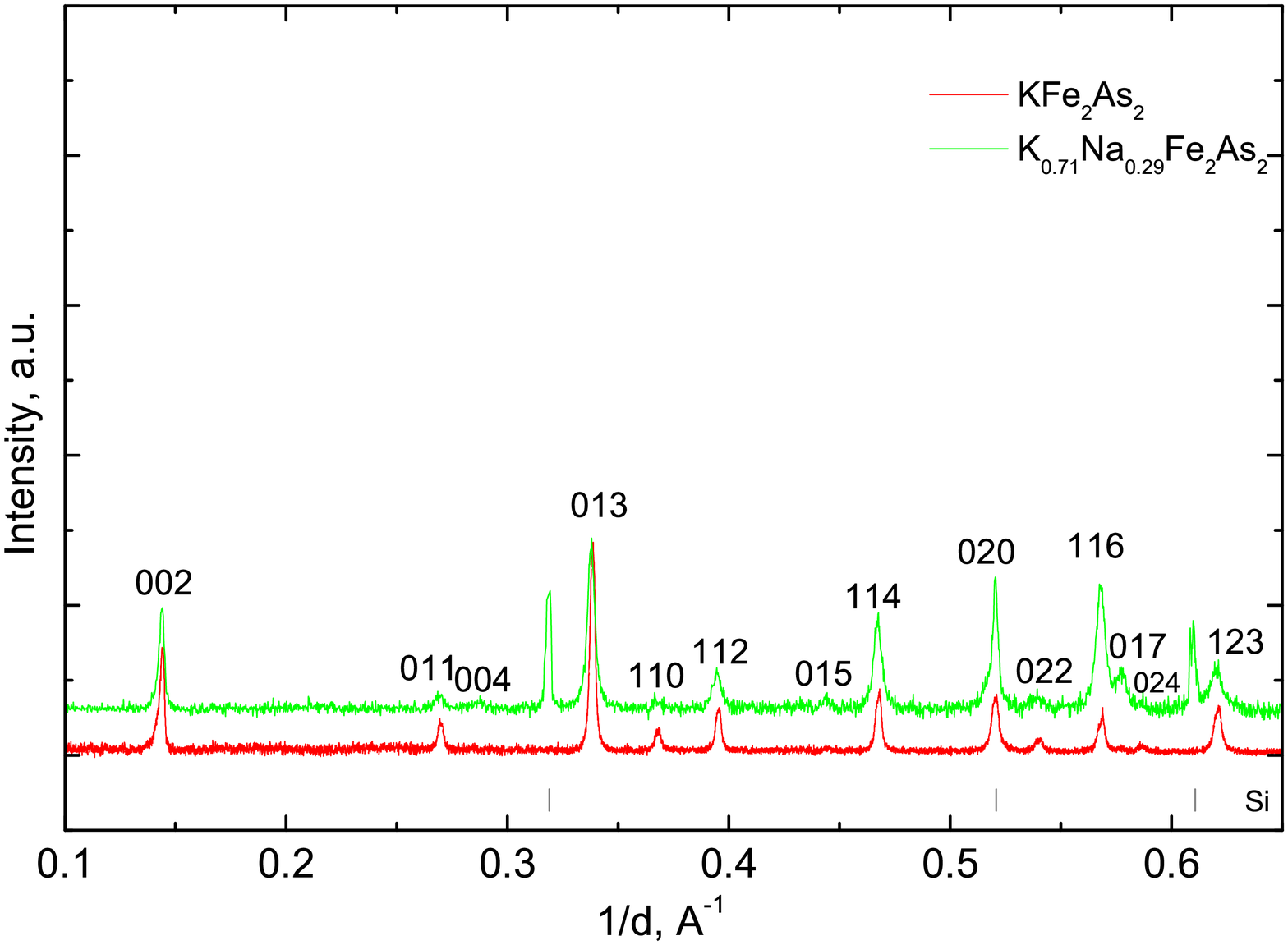}
\caption{The powder x-ray diffraction patterns for the stoichiometric KFe$_2$As$_2$ and Na doped samples.}
\label{XRD}
\end{figure}

\section*{x-ray analysis} 

To demonstrate the phase purity of the K$_{1-x}$Na$_x$Fe$_2$As$_2$ samples we 
performed powder x-ray diffraction measurements. The powders were obtained 
by milling the single crystals in an Ar atmosphere. The data 
for $x=0$ and $x=0.29$ are shown in Fig.\ \ref{XRD}. 
(In the latter case Si has been used as an internal standard.)
It is seen that only 
reflections for the 122 phase are present in the data.


\begin{thebibliography}{100}
\bibitem{remark2} For the sake of completeness we mention that this
result is compatible with the mentioned above non-Fermi liquid theory 
for D = 2 proposed in Ref.\ \onlinecite{Zaanen2009} and the large mass
anisotropy of the order of 25 to 56 derived from the slopes of the upper critical field
at $T_c$ for the in-plane and out-off plane directions. 
Since the applicability
of the former is rather unclear at the moment, we will not discuss this 
case here.
\bibitem {remark}Strictly speaking, to the best of our knowledge,
 there is no quantitative and semi-analytical multiband pair-breaking theory or $T_c$-formula
available.
However, from the general behavior of multiband systems the following scenario
is expected. In relatively wide region where a quasi-linear decrease of $T_c$
is observed the experimental data can be reasonably well described by a single  
effective pair-breaking parameter which contains somehow averaged 
contributions from all bands, i.e.\ this region is expected to be described 
in the standard AG fashion as done here.
Deviations are expected in the lower-$T_c$ region, where isolated 
weakly coupled bands
would already quench. In the present case due to the always present
interband interaction $T_c$ vanishes for all subsystems at the same
$T_c$. As a consequence the contributions from the weaker bands continue
 to decay more slowly. The observation of such pecularities would be helpful to
 extract parameters of the weakly coupled bands. The presently 
 available experimental data cover only the upper "high"-$T_c$
 region and
 the applicability of an effective AG-approach seems to be justified.
\end{thebibliography}
\end{document}